\documentclass{article}

\usepackage{arxiv}

\usepackage{multirow}
\usepackage{color,soul}
\usepackage{subfigure}

\usepackage[utf8]{inputenc} % allow utf-8 input
\usepackage[T1]{fontenc}    % use 8-bit T1 fonts
\usepackage{hyperref}       % hyperlinks
\usepackage{url}            % simple URL typesetting
\usepackage{booktabs}       % professional-quality tables
\usepackage{amsfonts}       % blackboard math symbols
\usepackage{nicefrac}       % compact symbols for 1/2, etc.
\usepackage{microtype}      % microtypography
\usepackage{lipsum}
\usepackage{graphicx}
\graphicspath{ {./images/} }

\title{Social and Emotional Etiquette of Chatbots: A Qualitative Approach to Understanding User Needs and Expectations}

\author{
 Ekaterina Svikhnushina \\
  School of Computer and Communication Sciences\\
  EPFL\\
  Lausanne, Switzerland \\
  \texttt{ekaterina.svikhnushina@epfl.ch} \\
   \And
 Pearl Pu \\
  School of Computer and Communication Sciences\\
  EPFL\\
  Lausanne, Switzerland\\
  \texttt{pearl.pu@epfl.ch} \\
}

\begin{document}
\maketitle
\begin{abstract}
As chatbots are becoming increasingly popular, we often wonder what users perceive as natural and socially accepted manners of interacting with them. Some researchers maintain that humans should avoid engaging in emotional conversations with chatbots, while others have started building empathetic chatting machines using the latest deep learning techniques. To understand if chatbots should comprehend and display emotions, we conducted semi-structured interviews with 18 participants. Our analysis revealed their overall enthusiasm towards emotionally aware agents. More importantly, users’ intention to accept emotional chatbots seem to hinge on how these agents respond to our specific emotions, rather than just the ability to detect human emotions. Our findings also disclosed the specific application domains where emotionally intelligent technology could improve user experience. To conclude, we summarized a set of emotion interaction patterns that inspire users’ intention to adopt such technology as well as guidelines useful for the development of emotionally intelligent chatbots.
\end{abstract}

% keywords can be removed

\section{Introduction}
The use of voice and chat-based conversational agents is on the rise. Facebook has recently announced that the number of bots on Messenger Platform exceeded 300,000 \cite{boiteux2018messenger}. Gartner forecasts by the year 2022 chatbots will get involved in 85~\% of all customer service interactions \cite{bharaj2017gartner}. And a recent survey by PwC reported that over 700 US participants out of 1,000 use intelligent voice assistants, such as the Apple Siri, Amazon Alexa, or others to facilitate their everyday tasks \cite{pwc2018}. Hundreds of thousands of conversational agents are rapidly emerging in a diverse range of application domains.

\begin{figure}
    \centering
    \includegraphics[width=0.5\linewidth]{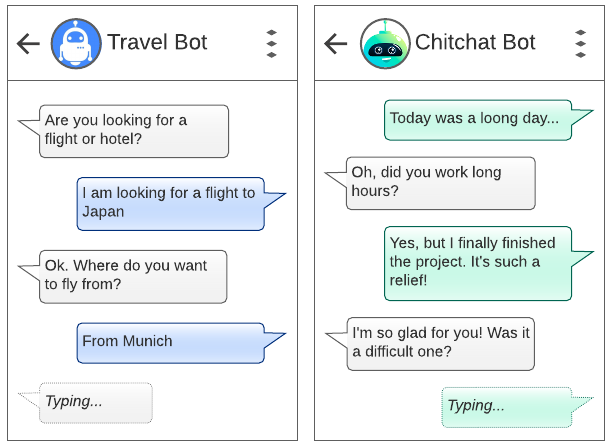}
    \caption{Example of dialogs with a task-oriented (left) and an open-domain (right) chatbots.}
    \label{fig:task-chat}
\end{figure}

Chatbots are Artificial Intelligence (AI) systems capable of producing natural language output based on human dialog input in the form of either text or speech. Sometimes they are called conversational agents \cite{Liao2018allwork}. There are mainly two types of chatbots (Figure \ref{fig:task-chat}): those that can accomplish specific tasks for users (task-oriented) and those that offer opportunities for users to socialize or be entertained, without necessarily focusing on a single particular topic (chitchat) \cite{gao2019convai}. While agents of the former type are currently the most widespread, to build a truly natural interaction with the latter type is among the most challenging tasks due to its open-domain nature \cite{Grudin2019chatbots}. While we do not specifically use the term ``open-domain chitchat'' in our discussions with users, we are primarily focused on the qualities of the second type.

Some researchers have investigated the perceptions, expectations, and concerns surrounding the use of conversational agents. Many previous studies pointed out the importance of naturalness, i.e. human-like qualities, in this technology \cite{Luger2016badpa, brandtzaeg2017people, Zamora2017sorry, thies2017how, Jain2018chatbots, neururer2018perceptions}.
The most frequently mentioned components comprising conversational agent's naturalness are: responding coherently with the preceding context \cite{Luger2016badpa, thies2017how, Jain2018chatbots, neururer2018perceptions}, anticipating user needs and questions \cite{Luger2016badpa, thies2017how}, understanding culture- or language-specific terms \cite{Zamora2017sorry, neururer2018perceptions}, facilitating input and response diversity \cite{Luger2016badpa, thies2017how, Jain2018chatbots, Muresan2019chats}, and developing a consistent personality \cite{thies2017how, Jain2018chatbots, neururer2018perceptions}. Although researchers emphasized the importance of integrating emotional intelligence \cite{brandtzaeg2017people, thies2017how, Zamora2017sorry}, no prior work has focused specifically on user expectations of the emotional qualities of chatbots. To address this gap, we have decided to conduct a qualitative user study as the first step towards understanding their expectations, especially concerning what emotional skills users would expect from artificial conversational agents.

We present the results of 18 semi-structured interviews with potential technology users. In the following, we first describe some background work in affect research, as well as surveying related qualitative studies of conversational agents. Further, we describe our study design, methodology, and main findings based on 400 affinity notes. We then discuss the implications of our work by identifying several design guidelines that can be useful for the development of emotionally aware chatbots.

\section{Background and Related Work}

\subsection{Development of Emotionally Aware Chatbots}

Affective computing, initiated by Picard \cite{picard2000affective} in the mid-1990s, is an essential aspect of human-computer interaction research. For example, earlier work showed that computer-initiated emotional support, such as demonstrating elements of active listening, empathy, and sympathy, can help users overcome frustration and manage negative emotional states \cite{klein2002computer}. Another study by Bickmore and Picard \cite{Bickmore2005} established that even after a long course of interaction, users found an embodied relational agent with deliberate social-emotional skills more respectful, appealing, and trustworthy than an equivalent task-oriented agent. Building on the previous experience of affective computing community, emotionally aware chatbots are equally believed to bring higher efficiency and engagement in human-computer interaction \cite{McDuff2018designing}. Recent progress in neural language modeling for response generation \cite{vinyals2015neural} has inspired an expanding number of papers focusing on introducing emotional awareness into neural network-based agents \cite{asghar2018affective, zhou2018emotional, zhou2018mojitalk, Hu2018touch, Huber2018emotional, Zhong2019affect, song2019generating, Xie2020}.  Several studies designed emotion-coping approaches by adjusting the neural network structure and the training objective function to make the model produce responses following a predefined strategy \cite{asghar2018affective, Zhong2019affect, Xie2020}. Other bodies of work employed explicit indicators, such as the use of emoji, image, or emotional category, to inform their model how to regulate the emotional response \cite{zhou2018emotional, zhou2018mojitalk, Hu2018touch, Huber2018emotional, song2019generating}. These papers mostly discuss the technical approaches to incorporate emotional intelligence into the chatbots. They do not explore what kind of emotional interaction is expected by the eventual technology users.

\subsection{Qualitative Studies of Chatbots}

Given the growing popularity of conversational agents, a number of studies analyzing user experience with this technology emerged. They mainly fall into two groups. Some researchers investigate user impressions of their interactions with available chatbots, while others deliberately focus on {\em future} interaction possibilities and elicit user expectations and concerns in those contexts.

The first type of work mostly develops around user reasons for interaction with existing agents and experience with them. Authors in \cite{Luger2016badpa} and \cite{brandtzaeg2017people} investigated the factors motivating and impeding user adoption of chatbots. Zamora \cite{Zamora2017sorry} explored these questions more specifically, focusing on user preference for input modality and domains of use. A number of works analyzed current user experience by conducting case studies of popular conversational agents \cite{Bentley2018, Porcheron2018, Muresan2019chats, purington_alexa_2017}. Jain et al. \cite{Jain2018chatbots} and Cowan et al. \cite{cowan_what_2017} focused on first-time and infrequent users of chatbots. Finally, Clark et al. \cite{clark_what_2019} researched on what characteristics are important for human conversation and how they apply to conversations with existing artificial agents. While several authors mentioned user interest in having more natural conversations with chatbots \cite{Zamora2017sorry, Jain2018chatbots, Muresan2019chats, purington_alexa_2017}, the majority of works mainly considered the pragmatic and operational characteristics of user-agent interaction, especially given the scarcity of open-domain chatbots on the market. Several works even questioned the appropriateness of human-like metaphor for conversational agents \cite{cowan_what_2017, Porcheron2018, clark_what_2019}. The supportive arguments for this view maintain that participants of their studies perceived social aspects of conversational interaction as an immaterial part of chatbot's performance \cite{clark_what_2019}, saw little need in agent's human-like behavior for it to address user tasks \cite{cowan_what_2017}, and did not treat it as a conversationalist \cite{Porcheron2018}. Authors of these works typically employ experience sampling, digital diary, and interviewing as methods for their studies.

Contrary to this practice, some researchers suggest that the way people perceive the technology in its current state and how they would prefer it to operate may considerably differ. Hence, they deliberately choose to focus on preferred user experience of future chatbots, to avoid any preconceptions caused by design peculiarities of existing agents. Authors in \cite{neururer2018perceptions} interviewed and surveyed researches from several relevant fields to determine characteristics of authenticity in future chatbots. Thies et al. \cite{thies2017how} ran an exploratory Wizard-of-Oz experiment to understand what chatbot personality traits would be preferred by their target audience, young urban Indians. Katayama et al. \cite{Katayama2019} conducted a mixed-method study to establish user expectations from an emotionally aware conversational agent on kinetic earable. So far this study is the closest to our work. They aimed at understanding the appropriate interaction patterns of situation-aware conversational agent in a wearable device. However, the scope of their study is limited to agent's speech prosody adaptation (such as pitch, rate, glottal tension), while ours is related to natural language generation.

As presented above, research focusing precisely on user expectations of emotional skills and social conversational principles for chatbots is limited. More precisely, it remains unclear what kind of emotional behaviors, or what we called social etiquette, will enable chatbots to comply better with user expectations.
In our work, we first decided to devote our efforts to a qualitative study aiming at investigating the interaction patterns that users would prefer an emotionally aware chatbot to follow and whether there are any concerns that could prevent them from adopting such systems.

\section{Method}

\subsection{Study Design}
We followed the semi-structure interview methodology to explore user expectations and concerns about emotionally aware chatbots. A detailed interview guide with open-ended questions was prepared to prompt the discussion with the participants. After developing the guide, we recruited the participants through the snowball sampling method \cite{Biernacki1981snowball}. In the invitation email, we provided a brief description of our research and interview procedure and informed the recipients about the incentives. Participants were recruited until saturation had occurred \cite{Guest2006}. We focused on participants who did not have a substantial exposure to chatbots before the interview to let them prospect the future technology as they would expect it to be and to avoid any biases resulting from previous (imperfect) experience. To make sure the participants could relate to the subject of discussion, we made sure that they used smartphones and computers on a regular basis and were familiar with messaging applications. Each participant was offered a small gift as a token of appreciation right after the interview, and two of them received smart speakers after a draw among all participants. All interviewees provided their consent for their data to be reported anonymously.

\subsection{Participants}
In total, 18 fluent or native English speakers (10 female, 8 male) from various backgrounds took part in our study. Over a half of the participants belonged to teenage (10--19 years old, 17~\%) and young adult (20--29 years old, 45~\%) age groups, with the remaining participants being almost equally distributed within four older age groups from 30--39 to 60--69 years old (38~\% in total). Most of the participants (67~\%) were nationals of European countries, and the others represented Asian, North- and South-American countries in roughly equal ratios.
All participants reported that they used computers and smartphones on a daily basis and were active users of different messaging platforms (e.g. WhatsApp, Telegram, WeChat) for communication with family, friends, and colleagues; their experience with chatbots was more narrow. All participants saw customer service chatbots online, but only a few (27.8~\%) mention using them before. Most participants (94.4~\%) indicated their familiarity with intelligent assistants (e.g. Siri, Google Assistant, Alexa, etc.), of this majority (55.6~\%) did not use them any more (as at the time of the interview) and others used them infrequently (38.8~\%). None of the participants had experience with a dedicated open-domain chatbot, nor was aware of their existence.

\subsection{Interviews}

Once agreed to take part in our study, the participants were asked to complete a basic demographic survey about their age group, nationality, and occupation. The following semi-structured interviews were organized either in-person (11 cases), or via Skype video-conference (7 cases). All of them took place between 2nd and 23rd October 2019, with each interview lasting about 40 minutes. All interviews but one were audio-recorded, with the participants' consent, and all were accompanied with hand-written notes either by the interviewer or interviewer's colleague.
Each interview covered four sections: 1) general questions about participant's background and experience with technology (computer, smartphone, messaging apps); 2) knowledge of and previous experience with chatbots; 3) qualities desired from the emotionally aware chatbots to make interaction with them more natural (if any), and purposes of such agents; 4) any concerns the participant might have about using such chatbots. Specifically, the first two parts were adjusted to make the participants speculate about their recent experience of social conversations (in person and via messaging apps) and interaction with chatbot technology respectively. This part helped the participants to draw the parallels between their human-human and human-machine communication experience and took about 15 minutes in each interview. In the following core parts of the sessions, the interviewees were spurred to reflect on what they expect from natural conversations with emotionally intelligent chatbots and what could make them feel restrained from using the system. These discussions lasted for 25 minutes on average.

\subsection{Data Analysis}

\begin{figure}[!b]
    \centering
    \includegraphics[width=0.5\linewidth]{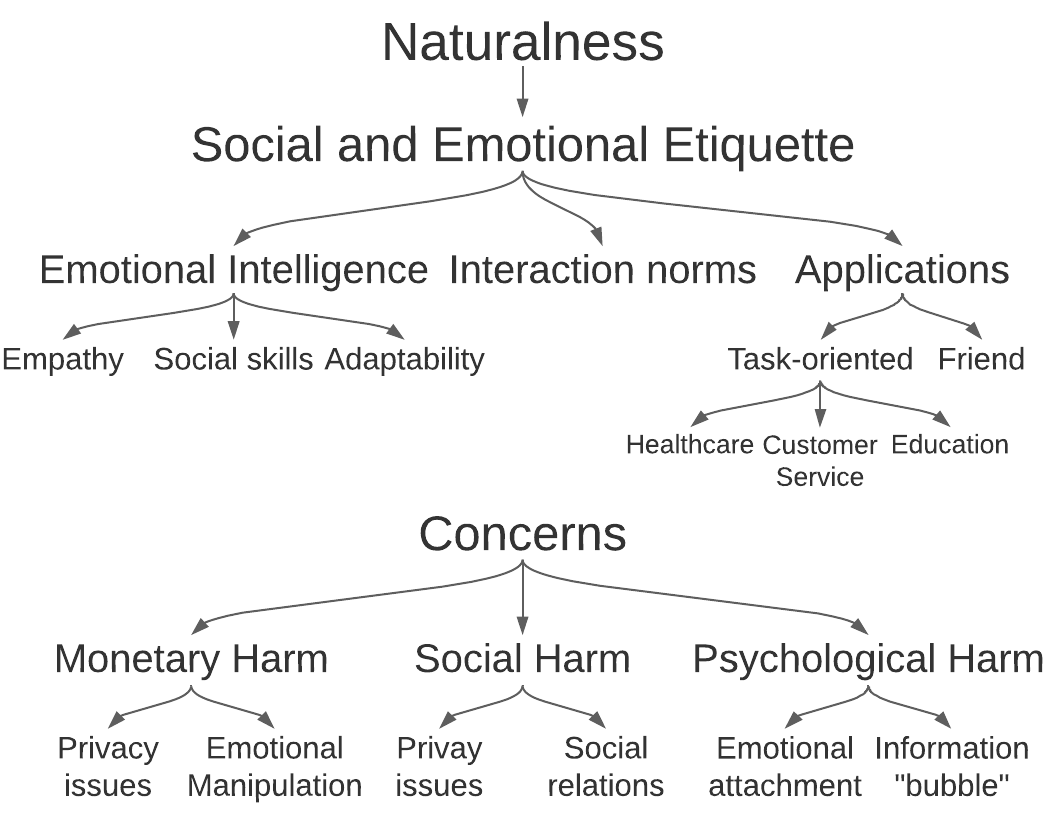}
    \caption{Structure of the affinity diagram.}
    \label{fig:affinity-wall}
\end{figure}

We used Affinity diagramming \cite{scupin1997kj} to analyze the interview content.
After each interview, the first author enriched the hand-written interview notes with missing comments and observations from audio records and extracted affinity notes with meaningful quotations from the participants and the researcher's remarks. If necessary, the interview guide was modified slightly before the next interview took place to ensure that all questions would be well understood by the participant. After all interviews were finished, several iterations of affinity diagramming took place. During the preliminary analysis, the first author clustered all resulting affinity notes according to emerging themes and validated the result with the second author. Three large themes describing chatbot's naturalness properties, participants' concerns, and application domains arose. The concept of emotion comprised a substantial part of naturalness and was also present in the other two themes. Overall, 400 affinity notes related to the concept of emotional awareness in chatbots, which accounted for over a half of all affinity notes in the initial diagram. Further, we examined these notes more closely. Specifically, the first author distributed emotion-related notes into sub-clusters and summarized their content with one representative topic. The sub-clusters, in turn, were grouped under top-level categories. The resulting affinity diagram, concerning emotional awareness of chatbots, was reviewed together with the second author and refined to reach its final version (Figure \ref{fig:affinity-wall}), which is discussed in the findings below.

\section{Findings}
\subsection{Expectations of Emotional Intelligence in Chatbots}
All participants of our study agreed that enabling more human-like behavior for the conversational agents could facilitate the interaction. Sixteen out of 18 interviewees expressed varying degrees of interest in chatbots with enhanced emotional capabilities. Seven participants felt highly enthusiastic about such agents, and the remaining 9 showed moderate excitement. Their expectations largely complied with an established notion of emotional intelligence, which includes: self-awareness, self-regulation, motivation, empathy, and social skills \cite{goleman1996emotional, goleman2009working}. As self-awareness and motivation rather refer to subjects that are endowed with consciousness, people attributed the other three qualities -- {\em empathy}, {\em social skills}, and {\em self-regulation} -- to their desired artificial conversational agents.

\subsubsection{Empathy}
Empathy is our ability to sense the feelings and emotions of others, take their perspective, understand their needs and concerns \cite{goleman2009working}.
When describing their expectations of chatbot's emotional behavior, the participants highlighted two main components: {\em recognition} of the speaker's emotional state and {\em expression} of emotion in accordance with the context. The principle desire was to feel understood by the chat agent and receive appropriate responses. As noted by U04:  \textit{``It needs to sound as if it has emotions, not only one emotion for all times. For example, it could be sad or happy or something like that: maybe, happy when you're happy and understanding when you're sad.''}

In addition to a straightforward way to treat the speaker's emotion by explicitly referencing the feeling (e.g. ``I see that you are frustrated.''), a number of other more subtle approaches were discussed during the interviews. Several participants mentioned interjections, \textit{``phases that people have in a usual talk, like "am", "ah", "seems to be", "you know\dots"''} (U09), as a way to express reactions, emotional states, and thought processes. Emojis and emoticons were also referenced a remarkable way of revealing emotion in chat. For example, U17 commented: \textit{``I use them sometimes to convey the atmosphere of "smiling conversation".''}

\subsubsection{Social skills}
Social skills concern the way how we manage relationships with others. These include a broad range of competencies from knowing how to communicate smoothly and managing conflicts to cooperating and bonding with people \cite{goleman2009working}. Speculating about their potential interaction, younger participants (below 30 years old) tended to be more open-minded about the social aspect of chatbots in everyday life. They enjoyed the idea of a conversational agent that could convey emotions during the dialog and presumed they would treat it as a friend. Interviewees felt excited about the possibility to engage with chatbots and share their feelings especially when they feel bored, lonely, or lacking motivation, as exemplified by the quote from U11: \textit{``Some people have only one person they are close to, so they might need another one. So for them, the [emotionally intelligent chatbot] would be very useful: not to feel alone and to actually feel like they are talking to someone and sharing something.''}

Meanwhile, both younger and older participants expressed interest in social skills for task-oriented chatbots.
From their perspective, it could improve their current experience in several domains by ensuring more appropriate responses and alleviating the embarrassment from talking to a new person.

\subsubsection{Self-regulation}
In relation to chatbots, the most frequently mentioned principles of self-regulation included trustworthiness and adaptability. The recurrent topic reflecting anticipated interaction development with the chatbot concerned \textit{``familiarity level''} (U06) with the user. Several participants commented that receiving overly positive replies from someone barely known would seem odd and awkward. Similar to relationship development with a newly met person, participants expected the chatbot to consider personal boundaries and gradually adjust to their style, motives, and language. Participant U06 pointed her concern about appropriate conversational style and importance of social chitchat for her: \textit{``Maybe it's different for my generation, but when I write an email or a message on WhatsApp, I always say `Bonjour \dots' and some greetings. I think this is quite important.''}
Participant U08 further supported the idea with another example from her personal experience: \textit{``I really like that some software, it tries to learn my language\dots it will predict what I would like to say in a way I personally say. So, it adapts to my style.''}

Depending on the participant's needs and attitude towards the natural language agents, some of them preferred the interaction to follow a more formal style. In contrast, others expected it to develop informally, similar to the way of communication with their friends. For example, U17 welcomed the idea to develop a more close relationship with the chatbot: \textit{``For me, it would be an amazing idea to have a kind of an online personal friend. So, you always share some thoughts with your friend, but this one can be both your diary and at the same time a psychologist who can always listen to you.''}
By comparison, U14 preferred more formal communication: \textit{``\dots sometimes I find the service may be too cold. But, for example, when I was in the US for a bit, it was extremely warm and welcoming, to the point that I found it intrusive. So, yeah, I'd say it should be polite and understanding the problem I'm facing.''}

\subsection{Chatbots in the Role of a Friend}
In our study, 10 out of 18 participants discussed the possibility to develop a friendship with a conversational agent in case it could demonstrate sufficient qualities of emotional awareness. They agreed that the chatbot should adjust to the user's emotional state, also taking its prior knowledge of the user into consideration, if possible. While it suggests a personalized approach, the participants concurrently described a number of emotional interaction patterns expected from the agent. The patterns mainly reflected the desired chatbot's responses to basic human emotions \cite{robinson2008brain}, such as happiness, sadness, or anger, and several more complex interactions. We summarize these expected patterns in Table \ref{tab:emotion} and consider them in greater detail below.

\begin{table}[!b]
    \centering
        \begin{tabular}{cc}
    \vspace{6pt} \\
           {\textbf{Input emotion}}
         & {\textbf{Response emotion}} \\
         \toprule
         \vspace{6pt}
           Happiness      &      Happiness     \\
         \vspace{4pt}
         \begin{tabular}[c]{@{}c@{}}Loneliness,\\ Sadness\end{tabular} & \begin{tabular}[c]{@{}c@{}}Compassion,\\ Interest\end{tabular} \\
         \vspace{4pt}
           Anger          & Disengagement        \\
         \vspace{4pt}
           Disappointment & Motivation        \\
         \vspace{4pt}
           Frustration    & \begin{tabular}[c]{@{}c@{}}Non-judging\\
                                               support\end{tabular} \\
         \bottomrule
    \end{tabular}
    \caption{Expected emotional interaction patterns described recurrently by the participants.
      }~\label{tab:emotion}
\end{table}

During the analysis, we observed that male participants tended to comment more on the playful and entertaining interaction aspects, while female interviewees mostly emphasized chatbot's supporting abilities. Overall, the participants expected it to share their joyful moments, \textit{``ask what happened''} (U17), and \textit{``be happy with them''} (U02). In times of trouble, when feeling lonely or sad, the participants would anticipate understanding and compassion from the chatbot. U02 summarized these expectations as follows: \textit{``I guess, if you're adding some excitement or frustration, then she [emotionally intelligent chatbot] should either be happy with you or try to make the voice more comforting.''} Importantly, our participants would like chatbots to \textit{``provide  feedback, but not just generic''} (U16).

In some cases, potential users would desire the conversational agent to express coaching and motivational qualities. According to them, chatbots should encourage users \textit{``to keep going''} (U07) both literally, promoting more physical activity and helping to establish a healthy lifestyle, and figuratively, supporting them when dealing with everyday problems. U05 would appreciate if a chatbot could assist him with behavior change: \textit{``It would be good if it acts as a coach who helps you avoid a bad habit or encourage you to exercise.''} Several other participants would like chatbots to ``educate users to manage their anger'' (U01): \textit{``Maybe for me, a bot should calm you down when you're angry. [It should] say, "Stop, I cannot talk with you like that. If you don't calm down, I will turn off."''} (U03). Turning to chatbots to get inspiration and reassurance was another recurrently discussed topic: \textit{``\dots if you have to spend long hours there, alone, doing some experiments, then it can make a conversation with you, cheer you up, look at your problems, maybe give some advice. It's a kind of a colleague that you might not have''} (U02).

Aligned with previous findings \cite{Zamora2017sorry, brandtzaeg2017people}, our participants expressed eagerness to share their frustration and negative thoughts with the chatbot due to the non-judging nature of such interaction. They found it appealing to have someone always available to validate their anxiety and stress without condemning the users. As spotted by U18: \textit{``If it’s very natural, it can also be in the consulting domain\dots Consulting --  sometimes emotionally, sometimes professionally, like therapy.''} Curiously, just having an empathetic listener to vent out was not sufficient.  From the participants' perspective, the crucial part of this interaction scenario was to receive some non-generic feedback from the chatbot, either advising the user how to overcome the problem or helping them to take their mind off by \textit{``starting another topic [for conversation]''} (U04).

\subsection{Emotionally Aware Chatbots in Targeted Domains}
According to previous works on task-oriented chatbots, users try to engage into social chitchat with them, even though these agents were originally designed to operate in a limited target domain \cite{Kopp2005max, Liao2018allwork, Yan2017}. For example, \cite{Yan2017} reported that nearly 80~\% of user utterances to the online shopping bot were chitchat queries. Since almost all of our study participants have had previous experience of interaction with this type of agents, they soon delved into discussing the emotional awareness of these chatbots. Many interviewees took positive attitude towards emotionally aware chatbots for customer service, health care, and educational domains. They expected that chatbots could potentially eliminate issues caused by human factors: computer agents are not subject to stress and tiredness and could always offer comforting advice to the client. In the case of customer service it could ensure \textit{``more natural and pleasant''} responses, so that \textit{``people would actually want to call customer service instead of googling their problem''} (U11). For medical advice, several participants anticipated responses from the chatbot to be more attentive than the ones from \textit{``an over-worked, over-stressed doctor''} (U15).

For the area of educational and professional training, several participants pointed out that conversational agents could make the services more available along with expressing higher involvement and interest into the tutoring sessions. Both for health care and educational domains some interviewees mentioned the clients might feel less embarrassed to share their questions with a chatbot than with an unknown person. For example, U14 mentioned: \textit{``I guess, for some medical issues people may be shy to actually talk to a real doctor\dots So, for this case it [emotionally intelligent chatbot] could be quite good''} (U14).

\subsection{Three Pillars of User Concerns}

In line with previous studies \cite{Luger2016badpa, Zamora2017sorry, cowan_what_2017}, the main factors causing user worries around conversational agents were uncertainty about trustworthiness and reliability of the system, as well as the risk of private information exposure. Chatbot's ability to treat emotions provoked several additional topics that disturbed our interview participants. During the analysis, we identified three major categories that describe user concerns about chatbots: \textit{monetary harm}, \textit{social harm}, and \textit{psychological harm}. 

\subsubsection{Monetary harm}
Predictably, financial damage primarily involved the risks around the participants' immediate personal means, such as bank accounts or social security numbers. People also felt apprehensive about the threat to employment opportunities in case the technology reaches sufficiently natural conversational abilities. Potential emotional awareness of chatbots further increased these concerns as people feared that for intruders, \textit{``it would be easier to influence you with emotion''} (U05).

\subsubsection{Social harm}
Concerns about the consequences for the social status of the users developed around the risks of sensitive information misuse by the chatbot operators. People questioned how the information they share with the agents would be stored and whether the third-parties could use it. They worried that in case of disclosure, some pieces of data might be used against themselves and cause social embarrassment. Participant U11 questioned: \textit{``What if it remembers something you shouldn't have said?''} Participant U14 further echoed her worry: \textit{``If there's anything linked to some kind of psychology, I would be very scared of what is being kept [by the chatbot], because in the future you can be considered unbalanced, or whatever.''}

Several participants also felt wary of the possibly addictive effect of highly human-like conversational agents. Similarly to the way how excessive smartphone usage negatively affects our social relations \cite{Genc2018}, they concerned that users might get too obsessed with flawless \textit{``virtual friends''} (U10) and isolate themselves from real human society. Participant U02 found this especially alarming for children: \textit{``I wouldn't want children to use this technology, for them not to get used to talking to a computer all the time instead of real people.''}

\subsubsection{Psychological harm}
Sometimes people develop an emotional attachment to objects and may experience anxiety and other negative emotions when facing a risk of losing these items \cite{yap2019unpacking}. Our participants mentioned that people would highly likely establish an affective connection with emotionally aware chatbots. In this case, a technical glitch or agent's discontinuation could cause strong user distress: \textit{``If some system or electricity failure happens, and the system gets reset, a person might not understand why it cannot remember anything anymore and feel very upset''} (U02).

Another thought-provoking point arose from people's experience with existing media resources. Several participants noted that some media adapts to personal interests of its users and focuses all the suggested content around them, possibly depriving the alternative views or unintentionally hiding \textit{``the best option''} (U14) from the user. It may deceive the users leading them to get trapped in \textit{``their bubble''} (U16), believing that everyone around adheres to the same beliefs. Some of our participants concerned that, given their anticipated personalization features, artificial conversational agents may further exacerbate this problem and cause psychological discomfort for the users. Participant U07 exemplified it with a personal anecdote: \textit{``I am also very worried \dots about the control the media has to shape my thinking, especially on Facebook. \dots It shows me posts that have the same point of view as other posts that I've read. I might read posts of some political area and then it will show me lots of similar posts. So, I might gradually start thinking that that's the only point of view.''}

\section{Discussion}

This study has investigated user expectations and concerns about interaction with emotionally aware chat agents and revealed a number of insights to consider for chatbot development. Below, we list 6 essential design implications resulting from our findings and further research opportunities.

\subsection{Design Implications}

\subsubsection{Endow chatbots with emotionality to enhance likability.}
Emotions form an essential part of human conversation. We use them in our daily chats with friends, family members, colleagues, retail assistants, and others. Emotional cues help us communicate our ideas clearer, share experiences, and form relationships. While several previous studies challenged the value of social skills for conversational agents \cite{cowan_what_2017, Porcheron2018, clark_what_2019}, our research findings demonstrated that users are eager to find emotional awareness in chatbots. For them, this is an essential part of natural interaction. Hence, enabling conversational agents with emotional intelligence can help designers ensure more pleasant user experience. This complies with the Media Equation theory \cite{reeves1996media}, suggesting that people apply rules and conventions of social human interaction to computers. According to Reeves and Nass, users appear considerably more positive about the computer system and their interaction experience with it if the computer exhibits human-like qualities, for example being polite, cooperative, or showing personality traits. People perceive such technology as friendlier and more supportive and feel more comfortable with it.

\subsubsection{Mirror positive, but carefully treat negative emotions.}
According to observations from social psychology, during communication people typically mimic each other’s emotional states \cite{Stevanovic2015}. In many technical papers focusing on the development of emotionally aware chatbots, the authors tend to employ this principle either by training their models to follow the interaction patterns existing in the human dialogs corpora \cite{zhou2018emotional} or even explicitly configuring the model to minimize the affect dissonance between the user input and generated response \cite{asghar2018affective}. Our study revealed that this mirroring approach is only partially valid. More specifically, users indeed expect chatbots to echo their positive emotions, for example, to share and promote user happiness. However, when experiencing negative feelings, people prefer the agent to act more intelligently than simply mirroring the speakers' emotions (Table \ref{tab:emotion}). Designers should enable the agents with abilities to demonstrate attention and meaningful support to help users overcome negative sentiments. People are more likely to engage with empathetic chatbots that can give personalized feedback to the users.

\subsubsection{Use implicit language markers to react to and express emotions.}
Chatbots designers should consider a variety of ways to enable chatbots to establish empathetic behavior by demonstrating and treating emotions. Participants of our study brought up two alternatives to expressing emotions through words: interjections and emojis. This agrees with previous studies, where authors showed that conversational agents employing interjections and filler words were perceived as more natural and engaging by the users \cite{marge2010towards, cohn2019large}. Emotive interjections \cite{Goddard2014} could enable chatbots to validate user emotions in a subtle and realistic manner, for example by saying \textit{Wow!} to express a positive surprise or \textit{Yikes!} to confirm their awareness of something bad and unexpected. Emojis and emoticons can also be employed by chatbots to express emotion and regulate the interaction, similarly to how people use them in computer-mediated communication between each other \cite{derks2008emoticons}.

\subsubsection{Align with user style and language.}
In human dialogs, people tend to converge on linguistic behavior and word choices to achieve successful and favorable communication \cite{Branigan2010, Pickering2004}. According to user expectations for chatbots' self-regulation, the same principle manifests for this type of human-computer interaction. Chatbot designers should ensure that the agents adapt to their users by learning their profile and utilizing preferred user vocabulary and conversational style. This is in line with several earlier findings \cite{Branigan2010, Thomas2018}, suggesting that linguistic alignment by computers with the users both in terms of word choices, i.e {\em what} things are said, and response style, i.e. {\em how} these things are said, can promote user positive feelings towards the computer and make communication more engaging.

\subsubsection{Maintain curiosity to sustain engagement.}
In a social conversation, curiosity provokes active listening and responding behavior, which represents a premise for pleasant interactions \cite{KASHDAN2006140, davis1982determinants}. Our study illustrated that people would value their interaction with chatbots that demonstrate interest and involvement into user preoccupations. By being curious about the user and proactively asking for clarifications, chatbots can prove their attentiveness and  understand better the issues faced by the user, ensuring more appropriate responses. At the same time, people themselves have a natural passion for learning and gaining new knowledge and understanding of things \cite{loewenstein1994psychology}. As reflected in our findings, the impossibility to understand how chatbots operate and treat the information shared by the users caused considerable concerns. Similarly, people worried that artificial agents would limit the informational content that might potentially interest the users or let them learn about alternatives. These issues could arise if chatbots fail to satisfy user curiosity. By providing convincing responses to user {\em why?} and {\em how?} questions, conversational agents can explain themselves better and enhance user trust of the technology.

\subsubsection{Be aware of user concerns: counterbalance with benevolence and privacy guarantees.}
The potential of endowing chatbots with social and emotional skills brings novel user concerns associated with this aspect. The participants of our study discussed probable social and psychological harms, which were not previously considered in the context of chatbots. Yet, privacy issues and financial risks constituted an outstanding fraction of user concerns, possibly because users already face these issues in the existing agents and can relate to them from experience \cite{Zamora2017sorry, cowan_what_2017, Huang2020}. To promote user trust of the technology, designers need to consider revealing the integrity of their chatbot \cite{mayer1995integrative}, i.e. its operational principles that would be acceptable for the users, such as transparent privacy policies and data security. Further, demonstrating that the chatbot is driven primarily by the user interests rather than other concealed motives could potentially offset user apprehension \cite{mayer1995integrative}.

\subsection{Limitations and Future Work}
This paper aimed at understanding how people envision their interaction with emotionally intelligent chat agents. We interviewed 18 participants of heterogeneous background aiming to acquire extensive response patterns. We did not compare cultural differences in the findings as the ratio of representatives of different cultures was relatively small in the sample. While we observed several gender- and age-related trends, given the scale of the study, neither can we claim any strong dependencies. Future work can investigate both cultural and gender- and age-specific questions closer by surveying a larger number of participants.

Our findings revealed the emotional interaction patterns most expected by the chatbot users. We propose further extending these results with more detailed studies. The community would benefit from a subtle analysis of expected interaction principles based on more fine-grained emotional categories. Given the established user desire for empathy in chatbots, one way to approach this problem could be to analyze the patterns existent in available corpora of empathetic human dialogs \cite{rashkin2019towards}. It may provide some insights into how emotional flow should develop through several conversational turns. Also, follow-up research could formalize the interaction rules more systematically by employing an established taxonomy of emotions, e.g. Plutchik's wheel of emotions \cite{plutchik2001nature}.

The subsequent technical work should focus on developing a conversational agent according to the established principles and evaluating actual user experience with it. As a part of our research, we have already started the development of an emotionally aware open-domain chatbot \cite{Xie2020}. In what follows, we plan to extend the scope of our developmental efforts to incorporate the findings from this qualitative study and then conduct an extensive user evaluation.

\section{Conclusion}
In our study, we took the first step towards a thorough understanding of user expectations and concerns about emotionally intelligent chat agents aiming at informing the community of chatbot developers and improving user experience with these agents. We employed the semi-structured interview method and deliberately elicited users’ future needs and preferences, especially concerning social and emotional etiquette that they believe to be desirable in their virtual counterpart. The findings demonstrated that most participants found chatbots empowered with emotional intelligence to be more likable, attentive, and pleasant to interact. Chatbots should demonstrate empathy and strong social skills to increase user engagement. Further investigation revealed a set of the most prominent emotional interaction principles with which users would like their chatbots to comply. Additional analysis also identified the application domains which users foresaw as the most beneficial for this technology, as well as pointing out the major factors causing their concerns around emotional agents. We summarized the insights from our study in a set of design guidelines and believe this knowledge is essential in shaping the future efforts of designers and developers of emotionally aware chatbots.

\section*{Acknowledgments}
This project has received funding from the Swiss National Science Foundation (Grant No. 200021\textunderscore184602). The authors also express gratitude to all the participants for sharing their ideas, and to Anuradha Welivita Kalpani, Kavous Salehzadeh Niksirat, and Yubo Xie for assisting the note-taking process during the interviews.

% unsrt

\bibliographystyle{ieeetr}  
\bibliography{references}  %%% Remove comment to use the external .bib file (using bibtex).


% George Kour and Raid Saabne.
% \newblock Real-time segmentation of on-line handwritten arabic script.
% \newblock In {\em Frontiers in Handwriting Recognition (ICFHR), 2014 14th
%   International Conference on}, pages 417--422. IEEE, 2014.

% George Kour and Raid Saabne.
% \newblock Fast classification of handwritten on-line arabic characters.
% \newblock In {\em Soft Computing and Pattern Recognition (SoCPaR), 2014 6th
%   International Conference of}, pages 312--318. IEEE, 2014.

% \bibitem{hadash2018estimate}
% Guy Hadash, Einat Kermany, Boaz Carmeli, Ofer Lavi, George Kour, and Alon
% \newblock Estimate and replace: A novel approach to integrating deep neural
%   networks with existing applications.
% \newblock {\em arXiv preprint arXiv:1804.09028}, 2018.

% \end{thebibliography}

\end{document}